\documentstyle[sprocl]{article}
\def\Journal#1#2#3#4{{#1} {\bf #2}, #3 (#4)}

\def\NPB{{\em Nucl. Phys.} B}
\def\PLB{{\em Phys. Lett.}  B}

\def\PRD{{\em Phys. Rev.} D}
\def\ZPC{{\em Z. Phys.} C}
\def\EPJC{{\em Eur. Phys. J.} C}
\def\IJMPA{{\em Int. J. of Mod. Phys.} A}
\def\PR{\em Phys. Rep.}

\def\be{\begin{equation}}
\def\ee{\end{equation}}
\def\bea{\begin{eqnarray}}
\def\eea{\end{eqnarray}}

\begin{document}

\hspace*{\fill}WUE-ITP-99-023

\hspace*{\fill}hep-ph/9909549

\vfill

\title{POLARIZATION AND SPIN EFFECTS IN NEUTRALINO PRODUCTION AND
  DECAY IN SUPERSYMMETRIC MODELS}

\author{ G. MOORTGAT-PICK \footnote{e-mail: gudi@physik.uni-wuerzburg.de},
S. HESSELBACH \footnote{e-mail: hesselb@physik.uni-wuerzburg.de},
F. FRANKE \footnote{e-mail: fabian@physik.uni-wuerzburg.de},
H. FRAAS \footnote{e-mail: fraas@physik.uni-wuerzburg.de} }

\address{ Institut f\"ur Theoretische Physik, Universit\"at
  W\"urzburg, Am Hubland, D-97074 W\"urzburg, Germany }


\maketitle\abstracts{
We study
polarization and 
spin correlation effects in the MSSM and in
extended supersymmetric models at an 
$e^+e^-$ linear collider with polarized beams.  
The production of light neutralinos
$e^+e^- \rightarrow \tilde\chi_1^0 \tilde\chi_2^0$
and the subsequent decay
$\tilde\chi_2^0 \rightarrow \tilde\chi_1^0 e^+e^-$ including full spin
correlations is
compared in three supersymmetric models: MSSM, NMSSM and an
$E_6$-inspired model with an additional $U(1)'$ gauge group.
It is shown that polarization and spin effects can lead to a complete
different behavior of these models. Finally, the dependence of the 
cross section and the decay angular distribution on the gaugino mass
parameter $M_1$ for polarized beams is briefly outlined.
}
  
\section{Introduction}
The production of neutralinos at 
an $e^+e^-$ linear collider 
and their subsequent decays offer
excellent opportunities to measure the
neutralino masses and mixings. 
Especially beam polarization and spin correlation effects play an
important role for determinating precisely 
the parameters of the underlying
supersymmetric model and 
discriminating between different
SUSY models.

In our contribution, we study the production of the light neutralinos 
$e^+e^- \rightarrow \tilde\chi_1^0 \tilde\chi_2^0$
with polarized beams and the subsequent
leptonic decay 
$\tilde\chi_2^0 \rightarrow \tilde\chi_1^0 e^+e^-$ in three different
SUSY models: the Minimal Supersymmetric Standard Model (MSSM)
with and without the GUT relation for the gaugino mass parameters
$M_1$ and $M_2$,
the Next-to-Minimal Supersymmetric Standard Model (NMSSM) 
with an additional Higgs 
singlet superfield and an $E_6$ inspired model with a new $U(1)'$ gauge boson.

Since the angular distribution of the decay products depend
on the neutralino polarization, the spin correlations between production
and decay are included. In fact, spin correlations turn out to have
a strong influence on the decay angular distribution which forbids simple
factorization into a production and a decay factor. 
For the MSSM the spin effects have been studied in Ref.~\cite{gudi}.
These methods are now applied to extended supersymmetric models with
a singlino-like lightest neutralino.

The models and the scenarios are described in the following section.
Numerical results for 
neutralino production cross sections, 
decay angular distributions and forward-backward-asymmetries
are presented in Secs.~3 and 4.   
We conclude this paper with
a short survey of the $M_1$-dependence of the cross sections and 
forward-backward asymmetries for polarized beams in the MSSM.

\section{Scenarios}

\subsection{MSSM}
In the MSSM \cite{mssm} we refer to the 
the DESY/EFCA reference scenario for the Linear Collider \cite{efca}
which is given
in Table 1. This scenario implies the GUT relation
$M_1/M_2=5/3 \tan^2 \theta_W$ and leads to gaugino-like light neutralinos with
masses $m_{\tilde{\chi}_1^0}= 72$ GeV and $m_{\tilde{\chi}_2^0}= 130$ GeV.
For comparison, these mass eigenvalues as well as the selectron masses 
$m_{\tilde{e}_L}= 197$ GeV, $m_{\tilde{e}_R}= 160$ GeV 
are fixed in all models.

\subsection{NMSSM}
The NMSSM \cite{nmssm} is the simplest extension of the MSSM by
a Higgs singlet field which enlarges the neutralino sector from four
neutralinos in the MSSM to five NMSSM neutralinos. 
New parameters in the neutralino sector are the singlet vacuum expectation
value $x$ and
the trilinear couplings $\lambda$ and $\kappa$ in
the superpotential.\cite{nmssmneutr} 
A scenario
with a singlino-like lightest neutralino (Table 1) leads to
significantly different signatures compared to the 
MSSM.\cite{franke} 
In this case the masses and
mixings of the neutralinos $\tilde{\chi}_{2,...,5}^0$ correspond to
$\tilde{\chi}_{1,...,4}^0$ of the MSSM with $\mu = \lambda x$.

\subsection{$E_6$-model}

Models with additional U(1) factors in the gauge group are
a further extension of the MSSM beyond the NMSSM. 
We study an $E_6$-model with one new gauge boson $Z'$ and an
extended Higgs sector with one singlet 
superfield \cite{e6model}, which contains 
six neutralinos.\cite{e6neutralino}
Assuming $M' = M_1$ for the U(1) gaugino mass
parameters the four lighter neutralinos have MSSM-like character.
For $M' \gg x$, however, the lightest neutralino can be a nearly pure
singlino.\cite{decarlosespinosa} 
Such a scenario
(Table 1) where the spectrum of the lighter neutralinos
is similar to the NMSSM will be discussed in the following.

\begin{table}[t]
\begin{center}
\begin{tabular}{|c|c|c|c|c|c|c|c|}
\hline
Model & $M_2/$GeV & $M'/$TeV & $\mu$/GeV & $x/$TeV & $\lambda$ & 
$\kappa$ & $\tan\beta$ \\ \hline
MSSM & 152 & - & 316 & - & - & -  & 3 \\ \hline
NMSSM & 262 & - & - & 1 & 0.9 & 0.0295 & 3 \\ \hline
$E_6$ & 270 & 22.3 & - & 3 & 0.15 & - & 3 \\ \hline
\end{tabular}
\end{center}
\caption{Parameters of the supersymmetric models. All
scenarios lead to neutralino masses
$m_{\tilde{\chi}_1^0}= 72$ GeV and $m_{\tilde{\chi}_2^0}= 130$ GeV.}
\end{table}

\section{Cross sections}
Figs. 1--3 show the cross sections for neutralino production for
unpolarized beams and for the different polarization configurations.
In our scenarios with gaugino- and singlino-like light neutralinos,
they mainly depend on the selectron masses and on the 
following combinations of the 
$e\tilde{e}\tilde{\chi}_i^0$-couplings $f_i^L$ and $f_i^R$
of the left- and right-handed
selectrons: \cite{mssm}
\bea
\mbox{MSSM}: & f^L_1 f^L_2=-0.2, & f^R_1 f^R_2=-0.12,\\
\mbox{NMSSM}: & f^L_1 f^L_2=-0.0007, & f^R_1 f^R_2=0.06, \\
\mbox{$E_6$}: & f^L_1 f^L_2=-0.003, & f^R_1 f^R_2=-0.03. 
\eea

In the MSSM, the unpolarized
$\tilde{\chi}_1^0\tilde{\chi}_2^0$ cross 
section reaches its maximum value of $\sim 90$ fb
at about threshold and decreases for
CMS-energies of 2 TeV to 10 -- 20 fb.
Due to the dominating left-handed selectron couplings
the beam polarizations $(-+)$ 
(left handed
polarized electrons and right-handed polarized positrons) 
enhance the cross sections by a factor of about 2. 

In the extended models 
(Figs.~2, 3), the $\tilde{\chi}_1^0\tilde{\chi}_2^0$ cross sections
are generally small ($\sim 1$ -- 10 fb) due to the weak couplings 
of the singlino-like 
$\tilde{\chi}_1^0$. Nevertheless, even a neutralino with a 
$ 99\%$ singlino component can be directly produced at a
linear collider with a high luminosity $L=500$ fb${}^{-1}$ which
corresponds to the TESLA scheme.\cite{richard}
Cross sections for the NMSSM and $E_6$-model are rather 
similar with the exception of the 
$Z'$ peak at $\sqrt{s}=1264$ GeV. Since the additional gauge boson
considerably couples to the singlino-like $\tilde{\chi}_1^0$ this peak is 
rather distinct for $\tilde{\chi}_1^0\tilde{\chi}_2^0$ production.

Contrary to the MSSM scenario 
the cross sections are enhanced for the 
polarization configuration $(+-)$ by 
a factor 3 
due to the dominating right-handed
couplings, while they are strongly suppressed for the opposite beam
polarization.

\begin{figure}
\begin{minipage}[p]{6cm}
\begin{picture}(6,4)
\put(0,0){\includegraphics{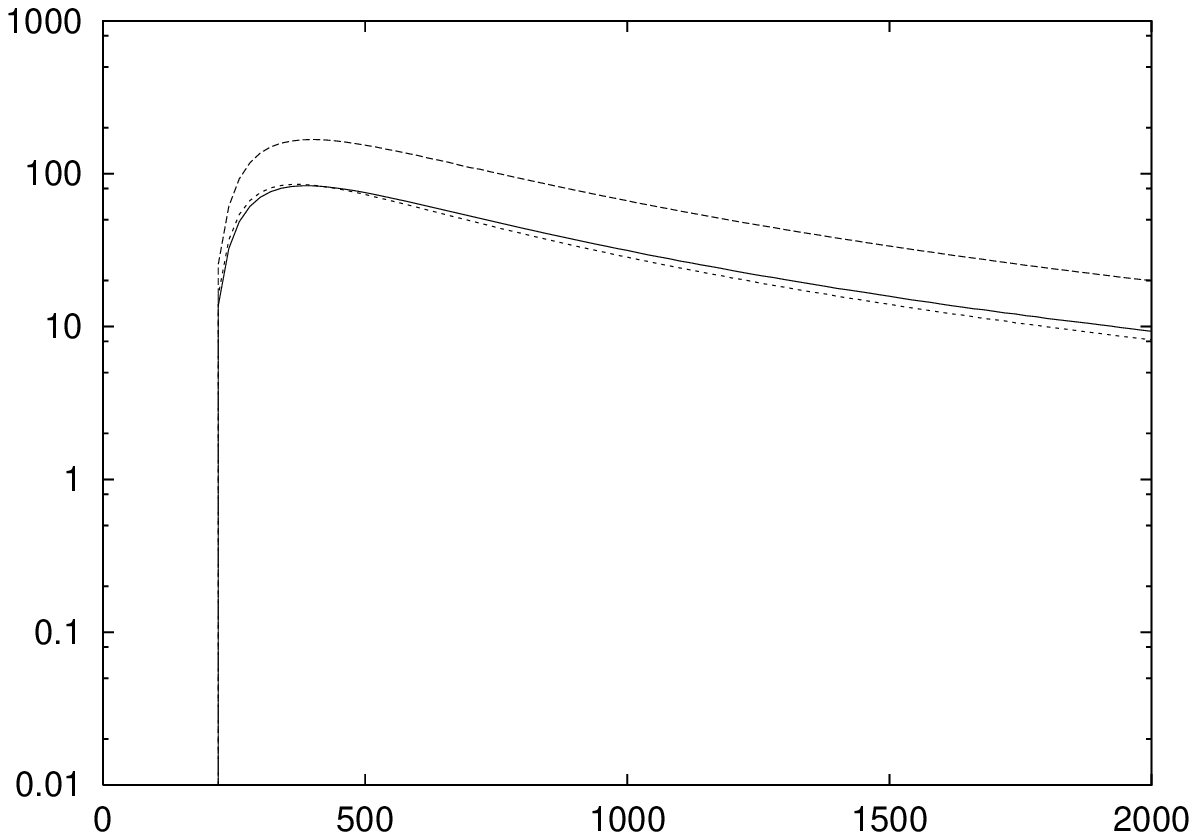}}
\put(5.2,-.4){\tiny $ \scriptscriptstyle \sqrt{s}$/GeV}
\put(-.5,3.5){$ \scriptscriptstyle\sigma_P/$\tiny fb}
\put(3.7,3.1){\tiny $(-+)$}
\put(3.7,2.1){\tiny $(+-)\approx (00)$}
\end{picture}\par\vspace{.5cm}
{\parbox{6cm}{\small Fig.~1: Cross section $\sigma_P(e^+ e^-\to
    \tilde{\chi}^0_1 \tilde{\chi}^0_2)$ for unpolarized beams $(00)$
and longitudinal beam polarizations $P_{-}=-85\%$, $P_{+}=+60\%$ $(-+)$ and
 beam polarizations $P_{-}=+85\%$, $P_{+}=-60\%$ $(+-)$ in the MSSM 
 (Table 1).}}
\end{minipage}\hfill
\begin{minipage}[p]{6cm}
\begin{picture}(6,4)
\put(0,0){\includegraphics{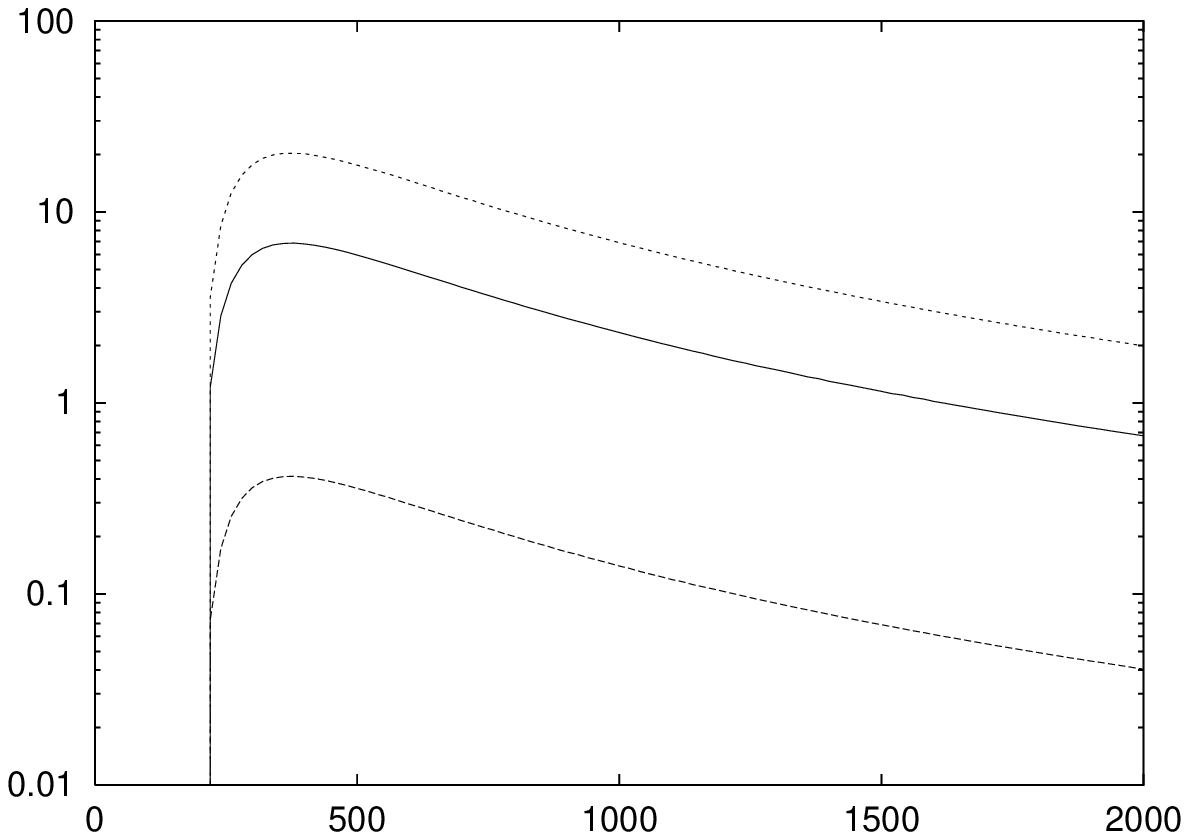}}
\put(5.2,-.4){\tiny $ \scriptscriptstyle \sqrt{s}$/GeV}
\put(-.5,3.5){\tiny $ \scriptscriptstyle\sigma_P/$fb}
\put(4,2.8){\tiny $(+-)$}
\put(4,2.25){\tiny $(00)$}
\put(4,1.1){\tiny $(-+)$}
\end{picture}\par\vspace{.5cm}
{\parbox{6cm}{\small Fig.~2: Cross section $\sigma_P(e^+ e^-\to
    \tilde{\chi}^0_1 \tilde{\chi}^0_2)$ for unpolarized beams $(00)$
and longitudinal beam polarizations $P_{-}=-85\%$, $P_{+}=+60\%$ $(-+)$ and
 beam polarizations $P_{-}=+85\%$, $P_{+}=-60\%$ $(+-)$ in the NMSSM
 (Table 1).}}
\end{minipage}
\end{figure}

\begin{figure}
\begin{minipage}[p]{6cm}
\begin{picture}(6,4)
\put(0,0){\includegraphics{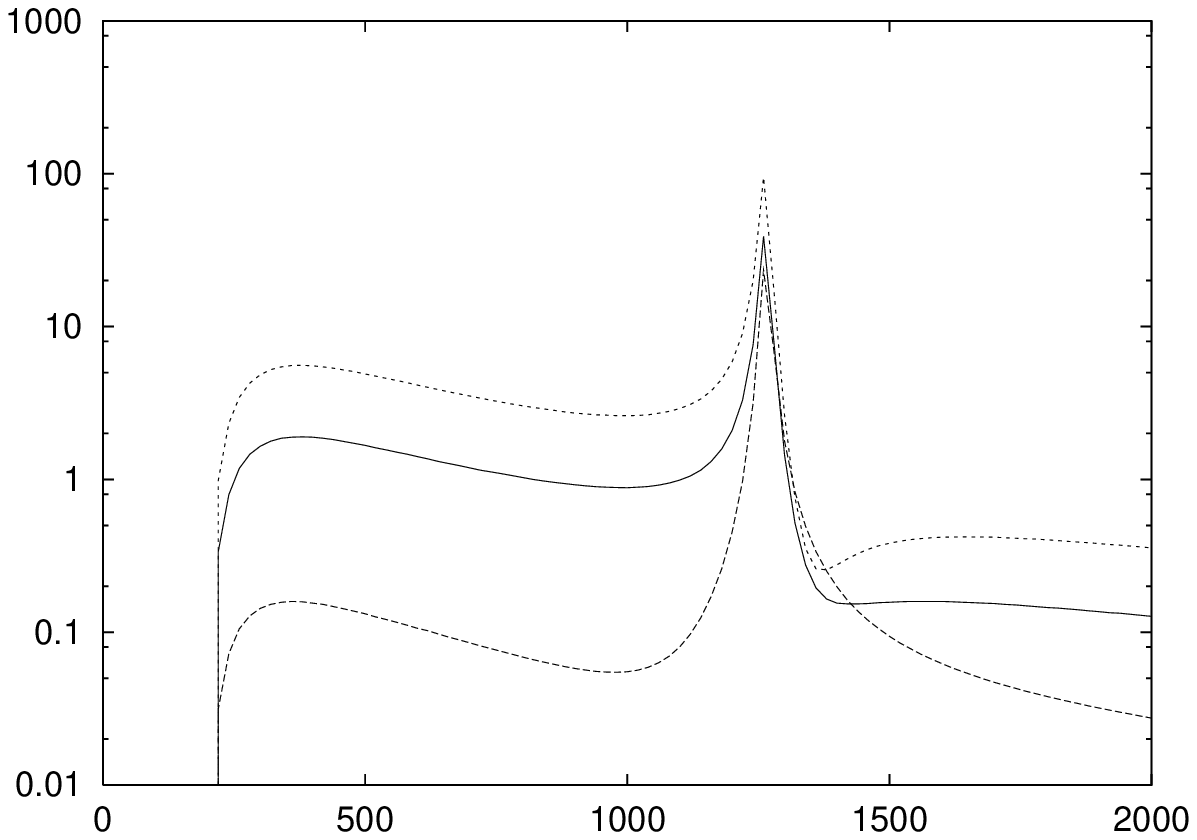}}
\put(5.2,-.4){\tiny $ \scriptscriptstyle \sqrt{s}$/GeV}
\put(-.5,3.5){\tiny $ \scriptscriptstyle\sigma_P/$fb}
\put(2,2.3){\tiny $(+-)$}
\put(2,1.8){\tiny $(00)$}
\put(2,1){\tiny $(-+)$}
\end{picture}\par\vspace{.5cm}
{\parbox{6cm}{\small Fig.~3:  Cross section $\sigma_P(e^+ e^-\to
    \tilde{\chi}^0_1 \tilde{\chi}^0_2)$ for unpolarized beams $(00)$
and longitudinal beam polarizations $P_{-}=-85\%$, $P_{+}=+60\%$ $(-+)$ and
 beam polarizations $P_{-}=+85\%$, $P_{+}=-60\%$ $(+-)$ in the $E_6$-model
 (Table 1). }}
\end{minipage}\hfill
\begin{minipage}[p]{6cm}
\begin{picture}(6,4)
\put(0,0){\includegraphics{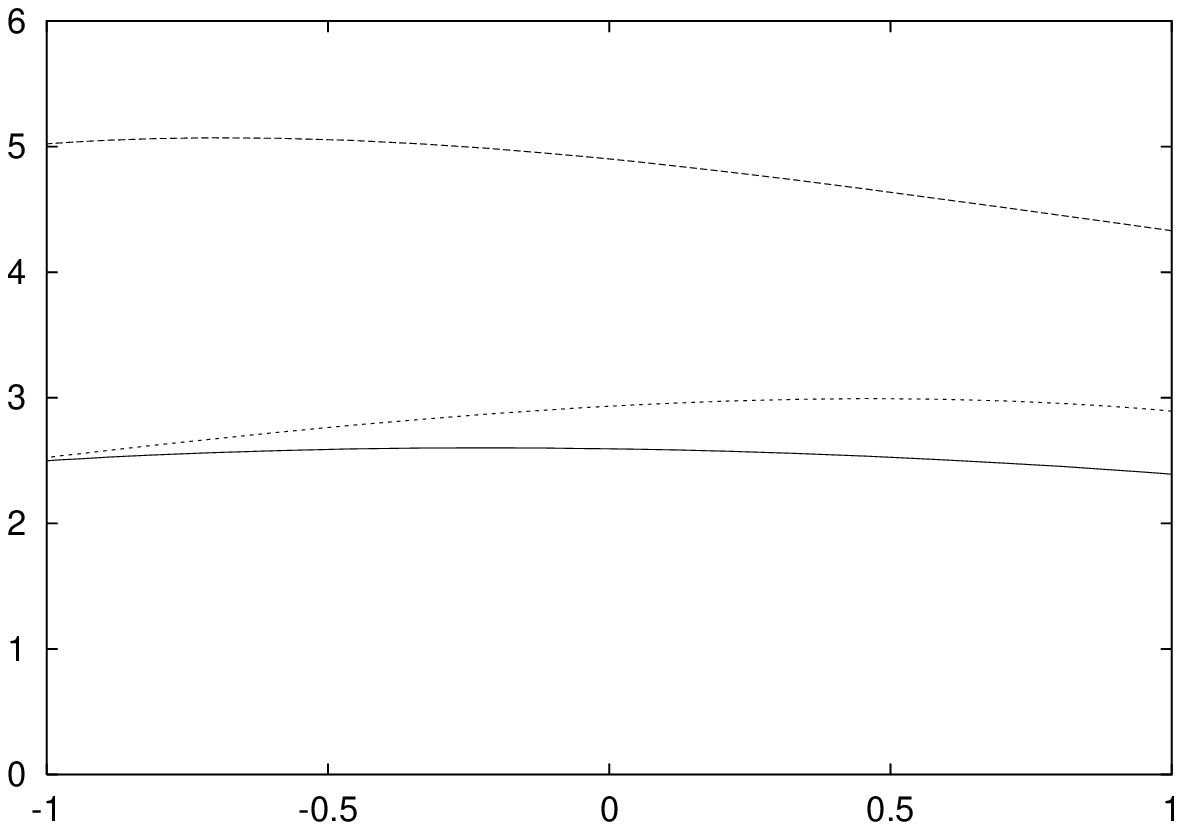}}
\put(5.2,-.2){$ \scriptscriptstyle \cos\Theta_{-}$}
\put(-.6,3.7){$ \scriptscriptstyle \frac{d\sigma_e}{d\cos\Theta_{-}}$}
\put(-.2,3.3){\tiny /fb}
\put(4,3.3){\tiny $(-+)$}
\put(4,2.25){\tiny $(+-)$}
\put(4,1.4){\tiny $(00)$}
\put(0.5,3.6){$ \scriptscriptstyle \sqrt{s}=m_{\tilde{\chi}^0_1}+
m_{\tilde{\chi}^0_2}+50$ \tiny GeV}
\end{picture}\par\vspace{.3cm}
{\parbox{6cm}{\small Fig.~4: Decay angular distribution of the decay
    electron in $e^+ e^- \to \tilde{\chi}^0_1 \tilde{\chi}^0_2\to
\tilde{\chi}^0_1 \tilde{\chi}^0_1 e^+ e^-$ with complete
spin correlations between production and decay for beam polarizations
$(00)$, $(-+)$ and $(+-)$ in the MSSM (Table 1).}}
\end{minipage}

\vspace{2cm}

\begin{minipage}[p]{6cm}
\begin{picture}(6,4)
\put(0,0){\includegraphics{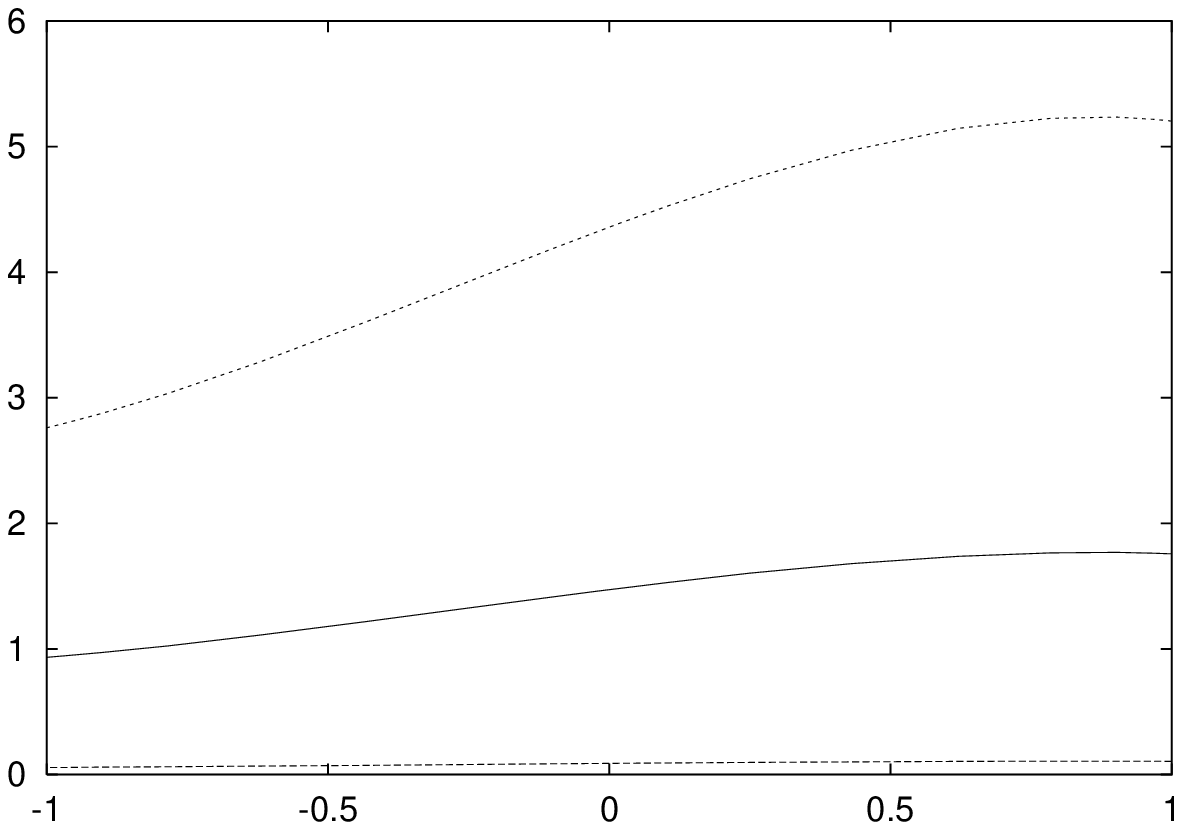}}
\put(5.2,-.2){$ \scriptscriptstyle\cos\Theta_{-}$}
\put(-.6,3.7){$ \scriptscriptstyle\frac{d\sigma_e}{d\cos\Theta_{-}}$}
\put(0.5,3.6){$ \scriptscriptstyle \sqrt{s}=m_{\tilde{\chi}^0_1}+
m_{\tilde{\chi}^0_2}+50$ \tiny GeV}
\put(-.2,3.3){\tiny /fb}
\put(4,3.5){\tiny $(+-)$}
\put(4,.3){\tiny $(-+)$}
\put(4,1.4){\tiny $(00)$}
\end{picture}\par\vspace{.3cm}
{\parbox{6cm}{\small Fig.~5: 
Decay angular distribution of the decay
electron in $e^+ e^- \to \tilde{\chi}^0_1 \tilde{\chi}^0_2\to
\tilde{\chi}^0_1 \tilde{\chi}^0_1 e^+ e^-$ with complete
spin correlations between production and decay 
for beam polarizations
$(00)$, $(-+)$ and $(+-)$ 
in the NMSSM (Table 1).}}
\end{minipage}\hfill
\begin{minipage}[p]{6cm}
\begin{picture}(6,4)
\put(0,0){\includegraphics{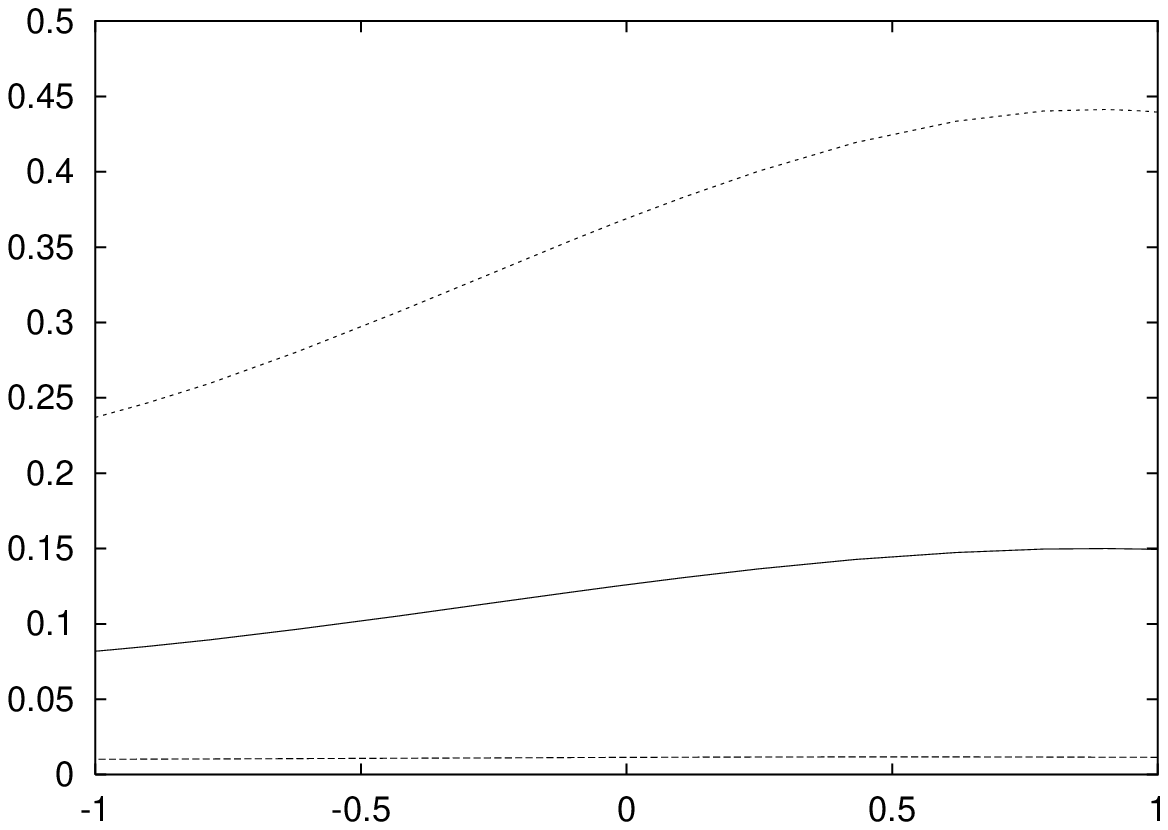}}
\put(5.2,-.2){$ \scriptscriptstyle \cos\Theta_{-}$}
\put(-.6,3.7){$ \scriptscriptstyle \frac{d\sigma_e}{d\cos\Theta_{-}}$}
\put(0.9,3.6){$ \scriptscriptstyle \sqrt{s}=m_{\tilde{\chi}^0_1}+
m_{\tilde{\chi}^0_2}+50$ \tiny GeV}
\put(-.2,3.3){\tiny /fb}
\put(4,3.5){\tiny $(+-)$}
\put(4,.3){\tiny $(-+)$}
\put(4,1.4){\tiny $(00)$}
\end{picture}\par\vspace{.3cm}
{\parbox{6cm}{\small Fig.~6: Decay angular distribution of the decay
electron in $e^+ e^- \to \tilde{\chi}^0_1 \tilde{\chi}^0_2\to
\tilde{\chi}^0_1 \tilde{\chi}^0_1 e^+ e^-$ with complete
spin correlations between production and decay 
for beam polarizations
$(00)$, $(-+)$ and $(+-)$ 
in the $E_6$-model (Table 1). }}
\end{minipage}
\end{figure}

\begin{figure}
\begin{minipage}[p]{6cm}
\begin{picture}(6,4)
\put(0,0){\includegraphics{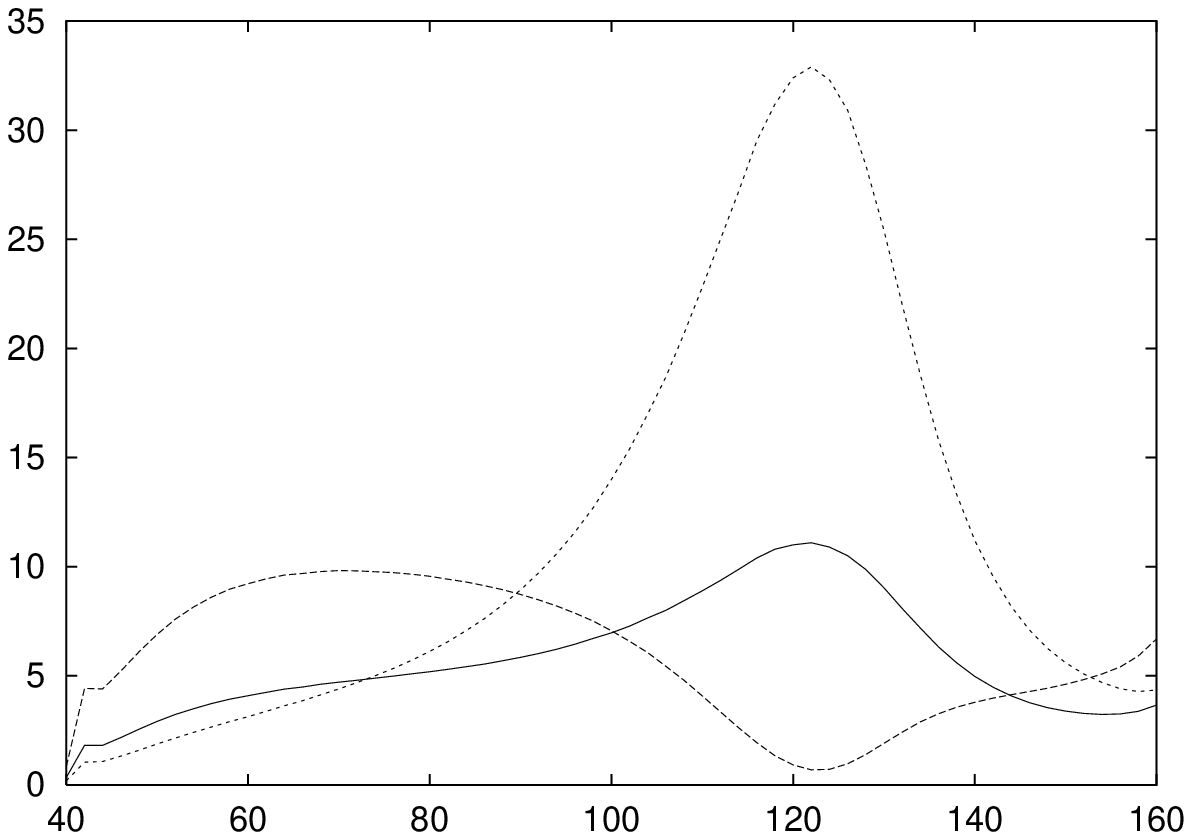}}
\put(5.2,-.4){\tiny $M_1$/GeV}
\put(-.5,3.5){\tiny $\sigma_e$/fb}
\put(3.8,2.9){\tiny $(+-)$}
\put(1.3,1.3){\tiny $(-+)$}
\put(3.9,1.4){\tiny $(00)$}
\put(0.5,3.6){$ \scriptscriptstyle \sqrt{s}=m_{\tilde{\chi}^0_1}+
m_{\tilde{\chi}^0_2}+50$ \tiny GeV}
\end{picture}\par\vspace{.5cm}
{\parbox{6cm}{\small Fig.~7: Cross section  $\sigma_e=\sigma_P(e^+ e^-\to 
    \tilde{\chi}^0_1 \tilde{\chi}^0_2)\times BR(\tilde{\chi}^0_2\to
    \tilde{\chi}^0_1 e^+ e^-)$
in dependence of the $M_1$-parameter 
for beam polarizations
$(00)$, $(-+)$ and $(+-)$ in the MSSM (Table 1).}}
\end{minipage}\hfill
\begin{minipage}[p]{6cm}
\begin{picture}(6,4)
\put(0,0){\includegraphics{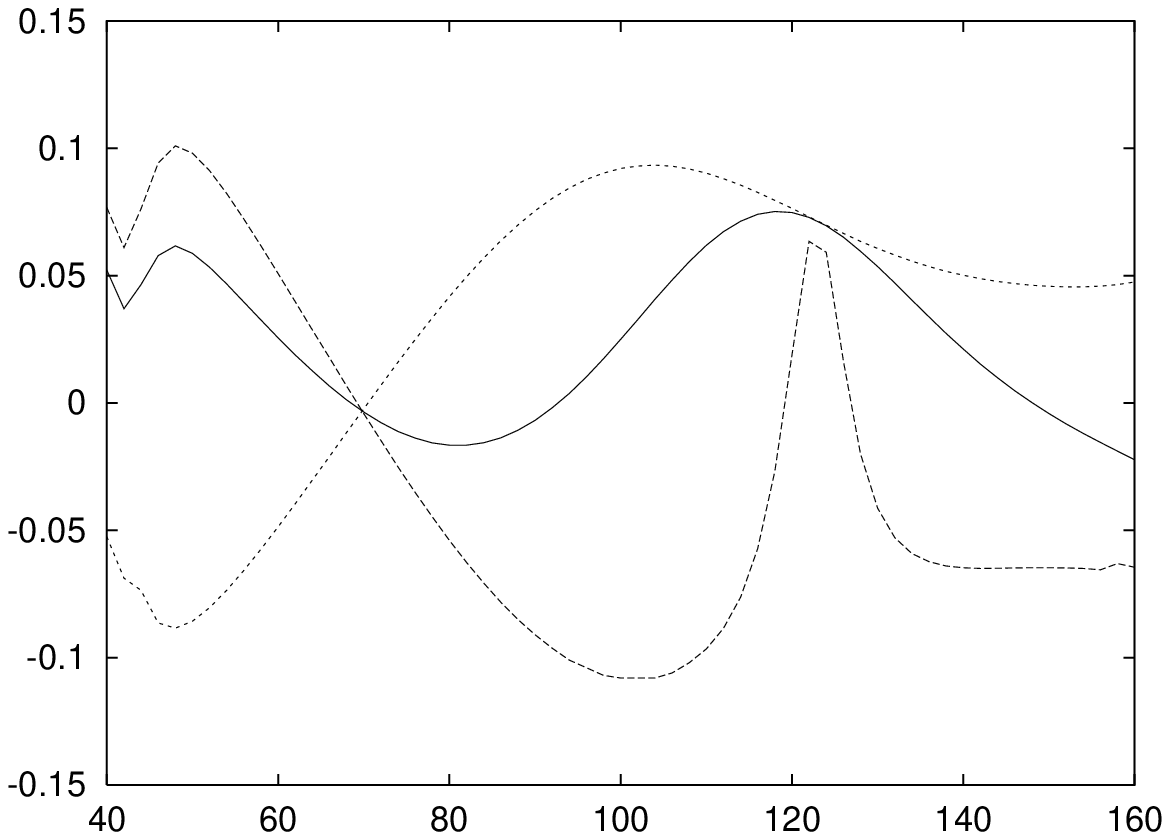}}
\put(5.2,-.4){\tiny $M_1$/GeV}
\put(-.5,3.5){\tiny $A_{FB}$}
\put(3,3.3){\tiny $(+-)$}
\put(3,.8){\tiny $(-+)$}
\put(2.3,1.9){\tiny $(00)$}
\put(0.9,3.6){$ \scriptscriptstyle \sqrt{s}=m_{\tilde{\chi}^0_1}+
m_{\tilde{\chi}^0_2}+50$ \tiny GeV}
\end{picture}\par\vspace{.5cm}
{\parbox{6cm}{\small Fig.~8: Forward-backward-asymmetry of lepton decay
    angular distribution in dependence of the $M_1$-parameter 
for beam polarizations
$(00)$, $(-+)$ and $(+-)$ in the MSSM (Table 1).}}
\end{minipage}
\end{figure}

\section{Decay angular distributions}

Since the spin correlations are strongest near threshold 
we study the decay angular distributions for the CMS-energy
$\sqrt{s}=m_{\tilde{\chi}^0_1}+m_{\tilde{\chi}^0_2}+50$~GeV.
For unpolarized beams the angular distribution between the incoming and the
decay electron in the MSSM (Fig.~4)
shows a very small forward-backward-asymmetry $A_{FB}^{(00)}\approx -1.2\%$
since the larger coupling of the left-handed selectron is
compensated by its bigger mass. With beam polarizations $(-+)$ 
$\tilde{e}_R$ exchange is heavier suppressed resulting in a more clearly 
negative
$A_{FB}^{(-+)}\approx -4.1\%$ while the opposite polarization configuration
leads to a positive $A_{FB}^{(+-)}\approx +3.8\%$ due to the suppressed
$\tilde{e}_L$ exchange.

In the NMSSM and $E_6$-scenarios with larger right-handed selectron
couplings, however, one obtains strong positive
decay angular asymmetries (Figs.~5, 6).
While the forward-backward-asymmetry of about $17\%$ in the NMSSM 
only weakly depends on the beam polarizations, the
spin correlation effects lead to
$A_{FB}^{(00)}\approx A_{FB}^{(+-)}\approx 16\%$ and 
$A_{FB}^{(-+)}\approx 4$\%
for the $E_6$-scenario.

\section{$M_1$-dependence}
Neglecting the GUT-relation between $M_1$ and $M_2$,
the changing selectron couplings lead to dramatically
different polarization effects which may be used for imposing bounds on  
$M_1$.\cite{M1} 
Fig.~7 shows the strong $M_1$-dependence of the cross section
$\sigma_e=\sigma_P(e^+ e^-\to \tilde{\chi}^0_1 \tilde{\chi}^0_2)\times
BR(\tilde{\chi}^0_2\to \tilde{\chi}^0_1 e^+ e^-)$ for polarized beams
in the MSSM.
The interplay of left- and right-handed selectron couplings now
causes a significant increase of 
the small forward-backward-asymmetry of Fig.~4; e.~g.~up 
to about $+ 5\%$ for unpolarized beams (Fig.~8). 

More details on the $M_1$-dependence can be found in Ref.~\cite{gudi1}.
The methods for the determination of the $M_1$ parameter may analogously
applied to extended models with similar results.

\section{Conclusion}
A high luminosity $e^+e^-$ linear collider is needed for the direct production
of a 
neutralino with a large singlino component in extended supersymmetric 
models
since the cross sections (beyond a possible $Z'$ peak) are highly suppressed 
compared to the MSSM.
Further, polarization of both beams is an
important tool to increase the event rates.

Polarization effects strongly depend on the
relation between left- and right-handed 
electron-selectron-neutralino-couplings and selectron masses. 
Therefore decay angular distributions and forward-backward-asymmetries
may help to distinguish between different supersymmetric models.

In a similar manner polarization effects reflect the dependence of the
neutralino couplings and masses on the gaugino mass parameter $M_1$ and 
supply additional informations for the determination of the
parameters of the underlying supersymmetric model.

\vspace{-1mm}
\section*{Acknowledgments}
\vspace{-1mm}

\begin{sloppypar}
We thank A.~Bartl and W.~Majerotto for valuable discussions.
G.M.-P.\ appreciates support from
DESY and the Institut f\"ur Hochenergiephysik der
\"Osterreichischen Akademie der Wissenschaften.
This work was supported by the Bundesministerium f\"ur Bildung
und Forschung, contract No.\ 05 7WZ91P (0) and by
the Fonds zur F\"orderung der wissenschaftlichen Forschung of Austria,
project No.\ P13139-PHY.
\end{sloppypar}


\vspace{-1mm}
\section*{References}
\vspace{-1mm}

\begin{thebibliography}{99}
\bibitem{gudi} G.~Moortgat-Pick, H.~Fraas,
\Journal{\PRD}{59}{015016}{1999}.

\bibitem{mssm} H.E. Haber, G.L. Kane \Journal{\PR}{117}{75}{1985}.  

\bibitem{efca} S. Ambrosanio, G.A. Blair, P. Zerwas, EFCA-DESY Linear Collider
Workshop,
{\em http://www.desy.de/conferences/ecfa-desy-lc98.html}.

\bibitem{nmssm} M. Drees, \Journal{\IJMPA}{4}{3635}{1989};
J. Ellis, J.F. Gunion, H.E. Haber, J. Roszkowski, F. Zwirner, 
\Journal{\PRD}{39}{844}{1989}. 

\bibitem{nmssmneutr} B.R. Kim, S.K. Oh, A. Stephan,
\Journal{\PLB}{336}{200}{1994};
F. Franke, H. Fraas, A. Bartl,
\Journal{\PLB}{336}{415}{1994}.

\bibitem{franke} F. Franke, H. Fraas, \Journal{\ZPC}{72}{309}{1996};
  U. Ellwanger, C. Hugonie, \Journal{\EPJC}{5}{723}{1998}.

\bibitem{e6model} J.F. Gunion, L. Roszkowski, H.E. Haber,
  \Journal{\PRD}{38}{105}{1988};
J.L. Hewett, T.G. Rizzo, \Journal{\PR}{183} {193}{1989};
  M.M. Boyce, M.A. Doncheski, H. K\"onig,
  \Journal{\PRD}{55}{68}{1997};
  M. Cveti\v{c}, D.A. Demir, J.R. Espinosa, L. Everett 
  and P. Langacker, \Journal{\PRD}{56}{2861}{1997};
  T. Gherghetta, T.A. Kaeding, G.L. Kane,
  \Journal{\PRD}{57}{3178}{1998}.

\bibitem{e6neutralino} J. Ellis, K. Enqvist, D.V. Nanopoulos,
  F. Zwirner, \Journal{\NPB}{276}{14}{1986};
  S. Nandi, \Journal{\PLB}{197}{144}{1987};
  E. Keith, E. Ma, \Journal{\PRD}{56}{7155}{1997}; 
  D. Suematsu, \Journal{\PRD}{57}{1738}{1998}.

\bibitem{decarlosespinosa} B. de Carlos, J.R. Espinosa,
  \Journal{\PLB}{407}{12}{1997}.

\bibitem{richard} F.~Richard, ECFA-DESY Linear Collider Workshop, Lund 1998. 

\bibitem{M1} 
             J.L. Kneur, G. Moultaka, \Journal{\PRD}{59}{015005}{1999};
             S.Y. Choi, A. Djouadi, H. Dreiner, J. Kalinowski,
             P. Zerwas, \Journal{\EPJC}{7}{123}{1999};
             S.Y. Choi, A. Djouadi, H.S. Song, 
             P. Zerwas, \Journal{\EPJC}{8}{669}{1999}.

\bibitem{gudi1} G.~Moortgat-Pick, H.~Fraas, A.~Bartl,
W.~Majerotto, \Journal{\EPJC}{9}{521}{1999}.

\end{thebibliography}
\end{document}